\documentclass[preprint,12pt]{elsarticle}
\usepackage{amssymb}
\usepackage{lineno}
\usepackage[tbtags]{amsmath}
\usepackage{graphicx,floatrow,threeparttable}

\begin{document}
\begin{frontmatter}
\title{Mechanical energy and mean equivalent viscous damping for SDOF fractional oscillators}
\author[rvt]{Jian Yuan\corref{cor}}
\cortext[cor]{Corresponding author. Tel.: +8613589862375.}
\ead{yuanjianscar@gmail.com}
\author{Bao Shi, Mingjiu Gai, Shujie Yang}
\address{Institute of System Science and Mathematics, Naval Aeronautical and Astronautical University, Yantai  264001, P.R.China}

\begin{abstract}
This paper addresses the total mechanical energy of a single degree of freedom fractional oscillator. Based on the energy storage and dissipation properties of the Caputo fractional derivatives, the expression for total mechanical energy in the single degree of freedom fractional oscillator is firstly presented. The energy regeneration due to the external exciting force and the energy loss due to the fractional damping force during the vibratory motion are analyzed. Furthermore, based on the mean energy dissipation of the fractional damping element in steady-state vibration, a new concept of mean equivalent viscous damping is suggested and the value of the damping coefficient is evaluated.
\end{abstract}


\end{frontmatter}

\section{Introduction}Viscoelastic materials and damping treatment techniques have been widely applied in structural vibration control engineering, such as aerospace industry, military industry, mechanical engineering, civil and architectural engineering [1]. Describing the constitutive relations for viscoelastic materials is a top priority to seek for the dynamics of the viscoelastically damped structure and to design vibration control systems.

Recently, the constitutive relations employing fractional derivatives which relate stress and strain in materials, also termed as fractional viscoelastic constitutive relations, have witnessed rapid development. They may be viewed as a natural generalization of the conventional constitutive relations involving integer order derivatives or integrals, and have been proven to be a powerful tool of describing the mechanical properties of the materials. Over the conventional integer order constitutive models, the fractional ones have vast superiority. The first attractive feature is that they are capable of fitting experimental results perfectly and describing mechanical properties accurately in both the frequency and time domain with only three to five empirical parameters [2]. The second is that they are not only consistent with the physical principles involved [3] and the molecule theory [4], but also represent the fading memory effect [2] and high energy dissipation capacity [5]. Finally, from mathematical perspectives the fractional constitutive equations and the resulting fractional differential equations of vibratory motion are compact and analytic [6].

Nowadays many types of fractional order constitutive relations have been established via a large number of experiments. The most frequently used models include the fractional Kelvin-Voigt model with three parameters [2]: $\sigma \left( t \right)={{b}_{0}}\varepsilon \left( t \right)+{{b}_{1}}{{D}^{\alpha }}\varepsilon \left( t \right)$, the fractional Zener model with four parameters [3]: $\sigma \left( t \right)+a{{D}^{\alpha }}\sigma \left( t \right)={{b}_{0}}\varepsilon \left( t \right)+{{b}_{1}}{{D}^{\alpha }}\varepsilon \left( t \right)$, and the fractional Pritz model with five parameters [7]: $\sigma \left( t \right)+a{{D}^{\alpha }}\sigma \left( t \right)={{b}_{0}}\varepsilon +{{b}_{1}}{{D}^{{{\alpha }_{1}}}}\varepsilon \left( t \right)+{{b}_{2}}{{D}^{{{\alpha }_{2}}}}\varepsilon \left( t \right)$.

Fractional oscillators, or fractionally damped structures, are systems where the viscoelastic damping forces in governing equations of motion are described by constitutive relations involving fractional order derivatives [8]. The differential equations of motion for the fractional oscillators are fractional differential equations. Researches on fractional oscillators are mainly concentrated on theoretical and numerical analysis of the vibration responses. Investigations on dynamical responses of SDOF linear and nonlinear fractional oscillators, MDOF fractional oscillators and infinite-DOF fractional oscillators have been reviewed in [8]. Asymptotically steady state behavior of fractional oscillators have been studied in [9, 10]. Based on the functional analytic approach, the criteria for the existence and the behavior of solutions have been obtained in [11-13], and particularly in which the impulsive response function for the linear SDOF fractional oscillator is derived. The asymptotically steady state response of fractional oscillators with more than one fractional derivatives have been analyzed in [14]. Considering the memory effect and prehistory of fractional oscillators, the history effect or initialization problems for fractionally damped vibration equations has been proposed by Fukunaga, M. [15-17] and Hartley, T.T., and Lorenzo, C.F. [18, 19].

Stability synthesis for nonlinear fractional differential equations have received extensive attention in the last five years. Mittag-Leffler stability theorems [20, 21] and the indirect Lyapunov approach [22] based on the frequency distributed model are two main techniques to analyze the stability of nonlinear systems, though there is controversy between the above two theories due to state space description and initial conditions for fractional systems [23]. In spite of the increasing interest in stability of fractional differential equations, there's little results on the stability of fractionally damped systems. For the reasons that Lyapunov functions are required to correspond to physical energy and that there exist fractional derivatives in the differential equations of motion for fractionally damped systems, it is a primary task to define the energies stored in fractional operators.

Fractional energy storage and dissipation properties of Riemann-Liouville fractional integrals is defined [24, 25] utilizing the infinite state approach. Based on the fractional energies, Lyapunov functions are proposed and stability conditions of fractional systems involving implicit fractional derivatives are derived respectively by the dissipation function [24, 25] and the energy balance approach [26, 27]. The energy storage properties of fractional integrator and differentiator in fractional circuit systems have been investigated in [28-30]. Particularly in [29], the fractional energy formulation by the infinite-state approach has been validated and the conventional pseudo-energy formulations based on pseudo state variables has been invalidated. Moreover, energy aspects of fractional damping forces described by the fractional derivative of displacement in mechanical elements have been considered in [31, 32], in which the effect on the energy input and energy return, as well as the history or initialization effect on energy response has been presented.

On the basis of the recently established fractional energy definitions for fractional operators, our main objective in this paper is to deal with the total mechanical energy of a single degree of freedom fractional oscillator. To this end, we firstly present the mechanical model and the differential equation of motion for the fractional oscillator. Then based on the energy storage and dissipation in fractional operators, we provide the expression of total mechanical energy in the single degree of freedom fractional oscillator. Furthermore, we analyze the energy regeneration due to the external exciting force and the energy loss due to the fractional damping force in the vibration processes. Finally, based on the mean energy dissipation of the fractional damping element in steady-state vibration, we propose a new concept of mean equivalent viscous damping and determine the expression of the damping coefficient.

The rest of the paper is organized as follows: Section 2 retrospect some basic definitions and lemmas about fractional calculus. Section 3 introduces the mechanical model and establishes the differential equation of motion for the single degree of freedom fractional oscillator. Section 4 provides the expression of total mechanical energy for the SDOF fractional oscillator and analyzes the energy regeneration and dissipation in the vibration processes. Section 5 suggests a new concept of mean equivalent viscous damping and evaluates the value of the damping coefficient.
Finally, the paper is concluded in section 6 with perspectives.

\section{Preliminaries}

\newdefinition{rmk}{Definition}
\begin{rmk}
 The Riemann-Liouville fractional integral for the function $f\left( t \right)$ is defined as
\begin{equation}
{}_{a}I_{t}^{\alpha }f\left( t \right)=\frac{1}{\Gamma \left( \alpha  \right)}\int_{a}^{t}{{{\left( t-\tau  \right)}^{\alpha -1}}f\left( \tau  \right)d\tau },
\end{equation}
where $\alpha \in {{R}^{+}}$ is an non-integer order of the factional integral, the subscripts $a$ and $t$ are lower and upper terminals respectively.
\end{rmk}
\begin{rmk}
The Caputo definition of fractional derivatives is
\begin{equation}
{}_{a}D_{t}^{\alpha }f\left( t \right)=\frac{1}{\Gamma \left( n-\alpha  \right)}\int_{a}^{t}{\frac{{{f}^{\left( n \right)}}\left( \tau  \right)d\tau }{{{\left( t-\tau  \right)}^{\alpha -n+1}}}},n-1<\alpha <n.
\end{equation}
\end{rmk}
\newtheorem{thm}{Lemma}
\begin{thm}
The frequency distributed model for the fractional integrator [33-35]
The input of the Riemann-Liouville integral is denoted by $v\left( t \right)$and output$x\left( t \right)$, then ${}_{a}I_{t}^{\alpha }v(t)$ is equivalent to
\begin{equation}
\begin{cases}
 \frac{\partial z(\omega ,t)}{\partial t}=-\omega z(\omega ,t)+v(t), \\
x(t)={}_{a}I_{t}^{\alpha }v\left( t \right)=\int_{0}^{+\infty }{{{\mu }_{\alpha }}(\omega )z(\omega ,t)d\omega }, \\
\end{cases}
\end{equation}
with ${{\mu }_{\alpha }}\left( \omega  \right)=\frac{\sin \left( \alpha \pi  \right)}{\pi }{{\omega }^{-\alpha }}$.

System (3) is the frequency distributed model for fractional integrator, which is also named as the diffusive representation.\\
\end{thm}
\begin{thm}
The following relation holds[26]
\begin{equation}\int_{0}^{\infty }{\frac{\omega {{\mu }_{\alpha }}\left( \omega  \right)}{{{\omega }^{2}}+{{\Omega }^{2}}}}d\omega =\frac{\sin \alpha \pi }{2{{\Omega }^{\alpha }}\sin \frac{\alpha \pi }{2}}.
\end{equation}
\end{thm}
\section{Differential equation of motion for the fractional oscillator}
This section will establish the differential equation of motion for a single degree of freedom fractional oscillator, which consists of a mass and a spring with one end fixed and the other side attached to the mass, depicted in Fig.1. The spring is a solid rod made of some viscoelastic material with the cross-sectional area $A$ and length $L$, and provides stiffness and damping for the oscillator.\\
\begin{figure}[H]
\centering
\includegraphics[scale=1.0]{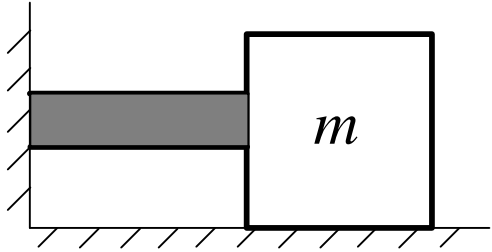}
\caption{Mechanical model for the SDOF fractional oscillator.}
\end{figure}In accordance with Newton's second law, the dynamical equation for the SDOF fractional oscillator is
\begin{equation}
m\ddot{x}\left( t \right)+{{f}_{d}}\left( t \right)=f\left( t \right), {{f}_{d}}\left( t \right)=A\sigma \left( t \right),
\end{equation}
where ${{f}_{d}}\left( t \right)$ is the force provided by the viscoelastic rod and can be separated into two parts: the resilience and the damping force. $f\left( t \right)$ is the vibration exciting force acted on the mass.\\
The kinematic relation is
\begin{equation}
\varepsilon \left( t \right)=\frac{x\left( t \right)}{L}.
\end{equation}
As for the constitutive equation of viscoelastic material, the following fractional Kelvin-Voigt model (7) with three parameters will be adopted
\begin{equation}
\sigma \left( t \right)={{b}_{0}}\varepsilon \left( t \right)+{{b}_{1}}D_{t}^{\alpha }\varepsilon \left( t \right),
\end{equation}
where $\alpha \in \left( 0,1 \right)$ is the order of fractional derivative, ${{b}_{0}}$ and ${{b}_{1}}$ are positive constant coefficients.\\
The above three relations (5) (6) and (7) form the following differential equation of motion for the single degree of freedom fractional oscillator
\begin{equation}
m\ddot{x}\left( t \right)+c{{D}^{\alpha }}x\left( t \right)+kx\left( t \right)=f\left( t \right),
\end{equation}
where $c=\frac{A{{b}_{1}}}{L}$, $k=\frac{A{{b}_{0}}}{L}$.\\
For the reason that the Caputo derivative is fully compatible with the classical theory of viscoelasticity on the basis of integral and differential constitutive equations [36], the adoption of the Caputo derivative appears to be the most suitable choice in the fractional oscillators. For the simplification of the notation, the Caputo fractional-order derivative ${}_{0}^{C}D_{t}^{\alpha }$is denoted as ${{D}^{\alpha }}$ in this paper.\\
Comparing the forms of differential equations for the fractional oscillator (8) with the following classical ones
\begin{equation}
m\ddot{x}\left( t \right)+c\dot{x}\left( t \right)+kx\left( t \right)=f\left( t \right),
\end{equation}
one can see that the fractional one (8) is the generalization of the classical one (9) by replacing the first order derivative $\dot{x}$with the fractional order derivative ${{D}^{\alpha }}x$. However, the generalization induces the following essential differences between them.

\begin{itemize}
  \item In view of the formalization of the mechanical model, the classical oscillator is composed of a mass, a spring and a dashpot, where $k$ is the stiffness coefficient of the spring offering restoring force $kx$ and $c$ is the damping coefficient of the dashpot offering the damping force $c\dot{x}$. The fractional oscillator is formed by a mass and an viscoelastic rod. The rod offers not only resilience but also damping force. In fractional differential equation(8), the coefficient $c$ and $k$ are determined by both the constitute equation (7) for the viscoelastic material and the geometrical parameters for the rod, which can be interpreted respectively as the fractional damping coefficient and the stiffness coefficient. As a result, the physical meaning of $c$ and $k$ in the fractional oscillator (8) and the classical one (9) are different. The fractional damping force can be viewed as a parallel of a spring component $kx\left( t \right)$ and a springpot component $c{{D}^{\alpha }}x\left( t \right)$ which is termed in [37] and illustrated in Fig.2. The hysteresis loop of the fractional damping force is dipicted in Fig.3.
\begin{figure}[H]
\centering
\includegraphics[scale=1.0]{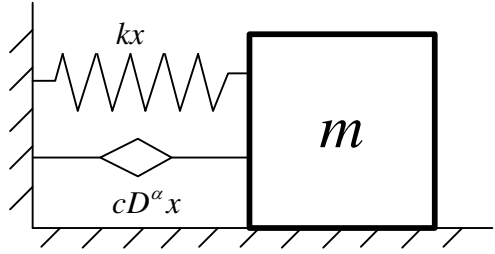}
\caption{Abstract mechanical model for the SDOF fractional oscillator.}
\end{figure}
\begin{figure}[H]
\centering
\includegraphics[scale=0.6]{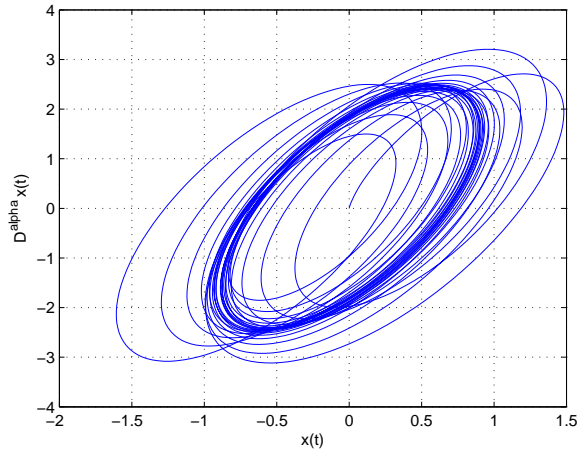}
\caption{Hysteresis loop of the fractional damping force.}
\end{figure}
  \item Fractional operators are characterized by non-locality and memory properties, so fractional oscillators (8) also exhibit memory effect and the vibration response is influenced by prehistory. While the classical one (9) has no memory effect and the vibration response is irrelevant with prehistory.
  \item In the aspect of mechanical energy, the fractional term ${{D}^{\alpha }}x$ in (8) not only stores potential energy but also consumes energy due to the fact that fractional operators exhibit energy storage and dissipation simultaneously [24]. As a result, the total mechanical energy in fractional oscillator consists of three parts: the kinetic energy $\frac{1}{2}m{{\dot{x}}^{2}}$ stored in the mass, the potential energy corresponding to the spring element $\frac{1}{2}k{{x}^{2}}$, and the potential energy $e\left( t \right)$ stored in the fractional derivative. However, in [38] the potential energy $e\left( t \right)$ stored in the fractional term ${{D}^{\alpha }}x$ has been neglected and the expression $\frac{1}{2}m{{\dot{x}}^{2}}+\frac{1}{2}k{{x}^{2}}$ for the total mechanical energy is incomplete.
\end{itemize}

\section{The total mechanical energy}
Given the above considerations, we present the total mechanical energy of the SDOF fractional oscillator (8)in this section. The fractional system is assumed to be at rest before exposed to the external excitation. We firstly analyze the energy stored in the Caputo derivative, based on which the expression for total mechanical energy is derived. Then we obtain the energy regeneration due to external excitation and the energy dissipation due to the fractional viscoelastic damping.\\
By definitions (1) and (2), the Caputo derivative is composed of one Riemann-Liouville fractional order integral and one integer order derivative,
\[{{D}^{\alpha }}x\left( t \right)={{I}^{1-\alpha }}\dot{x}\left( t \right).\]
In view of Lemma 1, the frequency distributed model for the Caputo derivative is
\begin{equation}
\begin{cases}
\frac{\partial z\left( \omega ,t \right)}{\partial t}=-\omega z\left( \omega ,t \right)+\dot{x}, \\
{{D}^{\alpha }}x\left( t \right)=\int_{0}^{\infty }{{{\mu }_{1-\alpha }}\left( \omega  \right)z\left( \omega ,t \right)d\omega }. \\
\end{cases}
\end{equation}
In terms of the fractional potential energy expression for the fractional integral operator in [24], the stored energy in the Caputo derivative is
\begin{equation}
e\left( t \right)=\frac{1}{2}\int_{0}^{\infty }{{{\mu }_{1-\alpha }}\left( \omega  \right){{z}^{2}}\left( \omega ,t \right)d\omega }.
\end{equation}
The total mechanical energy of the SDOF fractional oscillator is the sum of the kinetic energy of the mass $\frac{1}{2}m{{\dot{x}}^{2}}$, the potential energy corresponding to the spring element $\frac{1}{2}k{{x}^{2}}$, and the potential energy stored in the fractional derivative $ce\left( t \right)$
\begin{equation}
E\left( t \right)=\frac{1}{2}m{{\dot{x}}^{2}}+\frac{1}{2}k{{x}^{2}}+\frac{c}{2}\int_{0}^{\infty }{{{\mu }_{1-\alpha }}\left( \omega  \right){{z}^{2}}\left( \omega ,t \right)d\omega }.
\end{equation}
To analyze the energy consumption in the fractional viscoelastic oscillator, taking the first order time derivative of $E\left( t \right)$,one derives

\begin{equation}
\frac{dE\left( t \right)}{dt}=m\dot{x}\ddot{x}+kx\dot{x}+c\int_{0}^{\infty }{{{\mu }_{1-\alpha }}\left( \omega  \right)z\left( \omega ,t \right)\frac{\partial z\left( \omega ,t \right)}{\partial t}d\omega }.
\end{equation}
Substituting the first equation in the frequency distributed model (10) into the third term of the above equation (13), one derives
\begin{equation}
\begin{split}
\frac{dE\left( t \right)}{dt}&=m\dot{x}\ddot{x}+kx\dot{x}+c\int_{0}^{\infty }{{{\mu }_{1-\alpha }}\left( \omega  \right)z\left( \omega ,t \right)\left[ -\omega z\left( \omega ,t \right)+\dot{x} \right]d\omega }\\
& =m\dot{x}\ddot{x}+kx\dot{x}+c\dot{x}\int_{0}^{\infty }{{{\mu }_{1-\alpha }}\left( \omega  \right)z\left( \omega ,t \right)d\omega }\\
&\quad-c\int_{0}^{\infty }{\omega {{\mu }_{1-\alpha }}\left( \omega  \right){{z}^{2}}\left( \omega ,t \right)d\omega }.
\end{split}
\end{equation}
Substituting the second equation in the frequency distributed model (10) into the second term of the above equation (14), one derives

\begin{equation}
\begin{split}
\frac{dE\left( t \right)}{dt}&=m\dot{x}\ddot{x}+kx\dot{x}+c\dot{x}{{D}^{\alpha }}x-c\int_{0}^{\infty }{\omega {{\mu }_{1-\alpha }}\left( \omega  \right){{z}^{2}}\left( \omega ,t \right)d\omega }\\
&=\dot{x}\left[ m\ddot{x}+c{{D}^{\alpha }}x+kx \right]-c\int_{0}^{\infty }{\omega {{\mu }_{1-\alpha }}\left( \omega  \right){{z}^{2}}\left( \omega ,t \right)d\omega }.
\end{split}
\end{equation}
Substituting the differential equation of motion (8) for the fractional oscillator into the first term of the above equation (15), one derives
\begin{equation}
\frac{dE\left( t \right)}{dt}=f\left( t \right)\dot{x}\left( t \right)-c\int_{0}^{\infty }{\omega {{\mu }_{1-\alpha }}\left( \omega  \right){{z}^{2}}\left( \omega ,t \right)d\omega }.
\end{equation}
From Eq. (16) it is clear that the energy regeneration in the fractional oscillator due to the work done by the external excitation in unit time is
\begin{equation}
P\left( t \right)=f\left( t \right)\dot{x}\left( t \right).
\end{equation}
On the other hand, the energy consumption or the Joule losses due to the fractional viscoelastic damping is
\begin{equation}
J\left( t \right)=c\int_{0}^{\infty }{\omega {{\mu }_{1-\alpha }}\left( \omega  \right){{z}^{2}}\left( \omega ,t \right)d\omega }.
\end{equation}
The mechanical energy changes in the vibration process can be observed through the following numerical simulations. Parameters in the fractional oscillator (8) are taken respectively as $m=1$, $c=0.4$, $k=2$, $\alpha =0.56$, the external force are assumed to be $f\left( t \right)=30\cos 6t$. Fig.4 shows the fractional potential energy $ce\left( t \right)$; Fig.5 shows comparison between the fractional energy $ce\left( t \right)$and the total mechanical energy$E\left( t \right)$; Fig.6 illustrates the mechanical energy consumption $J\left( t \right)$.
\begin{figure}[H]
\centering
\includegraphics[scale=0.6]{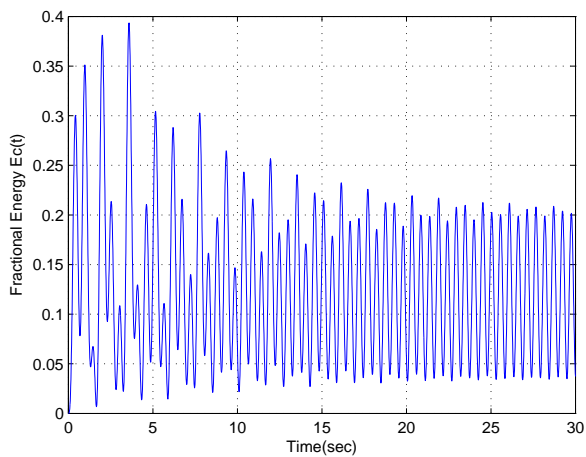}
\caption{Fractional energy of the SDOF fractional oscillator.}
\end{figure}
\begin{figure}[H]
\centering
\includegraphics[scale=0.6]{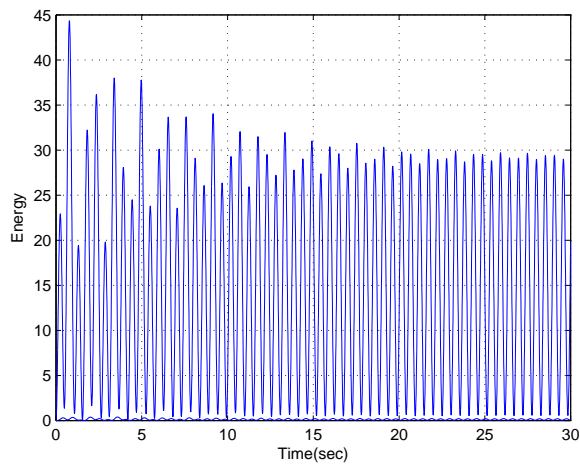}
\caption{Comparison between the fractional energy and the total mechanical energy.}
\end{figure}
\begin{figure}[H]
\centering
\includegraphics[scale=0.6]{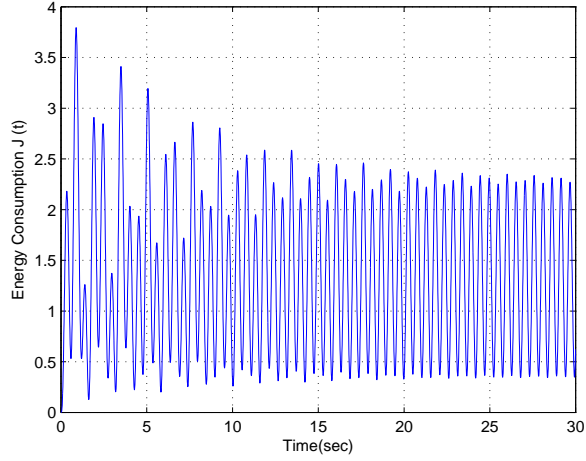}
\caption{Mechanical energy consumption in the SDOF fractional oscillator.}
\end{figure}
\newdefinition{rmka}{Remark}
\begin{rmka}
If the following modified fractional Kelvin-Voigt constitute equation (19)which is proposed in [39] is taken to describe the viscoelastic stress-strain relation
\begin{equation}
\sigma \left( t \right)={{b}_{0}}\varepsilon \left( t \right)+{{b}_{1}}{{D}^{{{\alpha }_{1}}}}\varepsilon \left( t \right)+{{b}_{2}}{{D}^{{{\alpha }_{2}}}}\varepsilon \left( t \right),
\end{equation}
with ${{\alpha }_{1}},{{\alpha }_{2}}\in \left( 0,1 \right)$, the differential equation of motion for the SDOF fractional oscillator is
\begin{equation}
m\ddot{x}\left( t \right)+{{c}_{1}}{{D}^{{{\alpha }_{1}}}}x\left( t \right)+{{c}_{2}}{{D}^{{{\alpha }_{2}}}}x\left( t \right)+kx\left( t \right)=f\left( t \right),
\end{equation}
where ${{c}_{1}}=\frac{A{{b}_{1}}}{L}$, ${{c}_{2}}=\frac{A{{b}_{2}}}{L}$, $k=\frac{A{{b}_{0}}}{L}$.

In view of the following equivalences (21) and (22) between the Caputo derivatives and the frequency distributed models
\begin{equation}
{{D}^{{{\alpha }_{1}}}}x\left( t \right)={{I}^{1-{{\alpha }_{1}}}}\dot{x}\left( t \right)\Leftrightarrow \begin{cases}
\frac{\partial {{z}_{1}}\left( \omega ,t \right)}{\partial t}=-\omega {{z}_{1}}\left( \omega ,t \right)+\dot{x}\left( t \right) \\
{{D}^{{{\alpha }_{1}}}}x\left( t \right)=\int_{0}^{\infty }{{{\mu }_{1-{{\alpha }_{1}}}}\left( \omega  \right){{z}_{1}}\left( \omega ,t \right)d\omega } \\
\end{cases}
\end{equation}
and
\begin{equation}
{{D}^{{{\alpha }_{2}}}}x\left( t \right)={{I}^{1-{{\alpha }_{2}}}}\dot{x}\left( t \right)\Leftrightarrow \begin{cases}
\frac{\partial {{z}_{2}}\left( \omega ,t \right)}{\partial t}=-\omega {{z}_{21}}\left( \omega ,t \right)+\dot{x}\left( t \right) \\
{{D}^{{{\alpha }_{2}}}}x\left( t \right)=\int_{0}^{\infty }{{{\mu }_{1-{{\alpha }_{2}}}}\left( \omega  \right){{z}_{2}}\left( \omega ,t \right)d\omega } \\
\end{cases}
\end{equation}
the total mechanical energy of the fractional oscillator (20) is expressed as
\begin{equation}
\begin{split}
E\left( t \right)&=\frac{1}{2}m{{\dot{x}}^{2}}+\frac{1}{2}k{{x}^{2}}+\frac{{{c}_{1}}}{2}\int_{0}^{\infty }{{{\mu }_{1-{{\alpha }_{1}}}}\left( \omega  \right)z_{1}^{2}\left( \omega ,t \right)d\omega }\\
& \quad+\frac{{{c}_{2}}}{2}\int_{0}^{\infty }{{{\mu }_{1-{{\alpha }_{2}}}}\left( \omega  \right)z_{2}^{2}\left( \omega ,t \right)d\omega }.
\end{split}
\end{equation}
In the above expression (23) for the total mechanical energy, $${{P}_{1}}\left( t \right)=\frac{{{c}_{1}}}{2}\int_{0}^{\infty }{{{\mu }_{1-{{\alpha }_{1}}}}\left( \omega  \right)z_{1}^{2}\left( \omega ,t \right)d\omega }$$ represents the potential energy stored in ${{D}^{{{\alpha }_{1}}}}x\left( t \right)$, whereas $${{P}_{2}}\left( t \right)=\frac{{{c}_{2}}}{2}\int_{0}^{\infty }{{{\mu }_{1-{{\alpha }_{2}}}}\left( \omega  \right)z_{2}^{2}\left( \omega ,t \right)d\omega }$$ represents the potential energy stored in ${{D}^{{{\alpha }_{2}}}}x\left( t \right)$.
Taking the first order time derivative of $E\left( t \right)$ in Eq.(23), one derives
\begin{equation*}
\begin{split}
\dot{E}\left( t \right)&=f\left( t \right)\dot{x}\left( t \right)-{{c}_{1}}\int_{0}^{\infty }{\omega {{\mu }_{1-{{\alpha }_{1}}}}\left( \omega  \right)z_{1}^{2}\left( \omega ,t \right)d\omega }\\
& \quad-{{c}_{2}}\int_{0}^{\infty }{\omega {{\mu }_{1-{{\alpha }_{2}}}}\left( \omega  \right)z_{2}^{2}\left( \omega ,t \right)d\omega }.
\end{split}
\end{equation*}
It is clear that the energy dissipation due to the fractional viscoelastic damping ${{c}_{1}}{{D}^{{{\alpha }_{1}}}}x$ is
\begin{equation}
{{J}_{{{\alpha }_{1}}}}\left( t \right)={{c}_{1}}\int_{0}^{\infty }{\omega {{\mu }_{1-{{\alpha }_{1}}}}\left( \omega  \right)z_{1}^{2}\left( \omega ,t \right)d\omega },
\end{equation}
and the energy dissipation due to the fractional viscoelastic damping ${{c}_{2}}{{D}^{{{\alpha }_{2}}}}x$ is
\begin{equation}
{{J}_{{{\alpha }_{2}}}}\left( t \right)={{c}_{2}}\int_{0}^{\infty }{\omega {{\mu }_{1-{{\alpha }_{2}}}}\left( \omega  \right)z_{2}^{2}\left( \omega ,t \right)d\omega }.
\end{equation}
\end{rmka}

\section{The mean equivalent viscous damping}
The resulting differential equations of motion for structures incorporating fractional viscoelastic constitutive relations to dampen vibratory motion are fractional differential equations, which are strange and intricately to tackled with for engineers. In engineering, complex descriptions for damping are usually approximately represented by equivalent viscous damping to simplify the theoretical analysis. Inspired by this idea, we suggest a new concept of mean equivalent viscous damping based on the expression of fractional energy (18). Using this method, fractional differential equations are transformed into classical ordinary differential equations by replacing the fractional damping with the mean equivalent viscous damping. The principle for the equivalency is that the mean energy dissipation due to the desired equivalent damping and the fractional viscoelastic damping are identical. \\
To begin with, some comparisons of the energy dissipation between the fractional oscillator (8) and the classical one (9) are made in the following.In view of the concept of work and energy in classical physics, the work done by any type of damping force is expressed as
\begin{equation}
W\left( t \right)=\int_{0}^{t}{{{f}_{c}}\left( \tau  \right)}dx\left( \tau  \right),
\end{equation}
where ${{f}_{c}}\left( t \right)$ is some type of damping force, $x\left( t \right)$ is the displacement of the mass.\\
In the classical oscillators, the viscous damping force is
$${{f}_{c1}}\left( t \right)=c\dot{x}\left( t \right).$$
The work done by the viscous damping force is
\begin{equation}
{{W}_{1}}\left( t \right)=\int_{0}^{t}{c\dot{x}\left( \tau  \right)}dx\left( \tau  \right)=\int_{0}^{t}{c{{{\dot{x}}}^{2}}\left( \tau  \right)}d\tau.
\end{equation}
It is well known that the energy consumption in unit time is
	
\begin{equation}
{{J}_{1}}\left( t \right)=c{{\dot{x}}^{2}}\left( t \right),
\end{equation}
which is equal to the rate of the work done by the viscous damping force
$${{J}_{1}}\left( t \right)=\frac{d{{W}_{1}}\left( t \right)}{dt}.$$
Obviously, the entire work done by the viscous damping force is converted to heat energy.\\
However, the case in the fractional oscillators is different. As a matter fact, the fractional damping force is
$${{f}_{c2}}\left( t \right)=c{{D}^{\alpha }}x\left( t \right).$$
The work done by the fractional damping force is
$${{W}_{2}}\left( t \right)=\int_{0}^{t}{c{{D}^{\alpha }}x\left( \tau  \right)}dx\left( \tau  \right).$$
Due to the property of energy storage and dissipation in fractional derivatives, the entire work done by the fractional damping force ${{W}_{2}}$ is converted to two types of energy: one of which is the heat energy
\[J\left( t \right)=c\int_{0}^{\infty }{\omega {{\mu }_{1-\alpha }}\left( \omega  \right){{z}^{2}}\left( \omega ,t \right)d\omega },\]
and the other is the fractional potential energy
\[P\left( t \right)=\frac{c}{2}\int_{0}^{\infty }{{{\mu }_{1-\alpha }}\left( \omega  \right){{z}^{2}}\left( \omega ,t \right)d\omega }.\]
However, in [40] the equivalent viscous damping coefficient was obtained by the equivalency
$$\oint{c{{D}^{\alpha }}x\left( \tau  \right)}dx\left( \tau  \right)=\oint{{{c}_{eq}}\dot{x}\left( \tau  \right)}dx\left( \tau  \right)$$
By this equivalency the properties of fractional derivative have been neglected and the work done by the fractional damping force is considered to be converted into the heat entirely. As a result, the above equivalency is problematic and the value of the derived equivalent viscous damping coefficient is larger than the actual value.
\\
In terms of the energy consumption (18), (24) and (25) due to the fractional damping force, we suggest a new the concept of mean equivalent viscous damping and evaluate the expression of the damping coefficient.
\\
Assuming the steady-state response of the fractional oscillator (8) is
\[x\left( t \right)=X{{e}^{j\Omega t}},\]
where $X$ is the amplitude and $\Omega$ is the vibration frequency.
\newcounter{Atte}
\newenvironment{Notes}{\begin{list}{\textbf{Step} \arabic{Atte}.}{%
\labelsep=1em\itemindent=3em%
\leftmargin=0pt%
\usecounter{Atte}}}{\end{list}}
\begin{Notes}
\item We firstly need to calculate the mean energy consumption due to the fractional viscoelastic damping element, i.e.
\begin{equation}
\overline{{{J}_{\alpha }}}\left( t \right)=c\int_{0}^{\infty }{\omega {{\mu }_{1-\alpha }}\left( \omega  \right)\overline{z{{\left( \omega ,t \right)}^{2}}}d\omega }.
\end{equation}	

To this end, we evaluate the mean square of $z\left( \omega ,t \right)$, i.e. $\overline{z{{\left( \omega ,t \right)}^{2}}}$.

In terms of the first equation in the diffusive representation of Caputo derivative (10)
\[\dot{z}\left( \omega ,t \right)=-\omega z\left( \omega ,t \right)+\dot{x}\left( t \right),\]
we get
\[z\left( \omega ,t \right)=\frac{\dot{x}\left( t \right)}{\omega +j\Omega }=\frac{j\Omega x{{e}^{j\Omega t}}}{\sqrt{{{\omega }^{2}}+{{\Omega }^{2}}}{{e}^{j\theta }}},\]
where $\theta =\arctan \frac{\Omega }{\omega }$.
\\
Furthermore we get
\begin{equation}
\overline{z{{\left( \omega ,t \right)}^{2}}}=\frac{1}{2}z\left( \omega ,t \right)z{{\left( \omega ,t \right)}^{*}}=\frac{1}{2}\frac{{{\Omega }^{2}}{{x}^{2}}}{{{\omega }^{2}}+{{\Omega }^{2}}},
\end{equation}
where $z{{\left( \omega ,t \right)}^{*}}$ is the complex conjugate of $z\left( \omega ,t \right)$.

Substituting Eq. (30) into Eq.(29), one derives

\begin{equation}
\begin{split}
\overline{{{J}_{\alpha }}}\left( t \right)&=c\int_{0}^{\infty }{\omega {{\mu }_{1-\alpha }}\left( \omega  \right)\overline{z{{\left( \omega ,t \right)}^{2}}}d\omega }\\
&=\frac{c}{2}{{\Omega }^{2}}{{X}^{2}}\int_{0}^{\infty }{\frac{\omega {{\mu }_{1-\alpha }}\left( \omega  \right)}{{{\omega }^{2}}+{{\Omega }^{2}}}}d\omega.
\end{split}
\end{equation}
Applying the relation (4) in Lemma 2 ,one derives
\begin{equation}
\int_{0}^{\infty }{\frac{\omega {{\mu }_{1-\alpha }}\left( \omega  \right)}{{{\omega }^{2}}+{{\Omega }^{2}}}}d\omega =\frac{\sin \left( 1-\alpha  \right)\pi }{2{{\Omega }^{\alpha }}\sin \left( \frac{1-\alpha }{2} \right)\pi }.
\end{equation}
Substituting Eq.(32) into Eq. (31) one derives
\begin{equation}
\overline{{{J}_{\alpha }}}\left( t \right)=\frac{c}{4}{{\Omega }^{1+\alpha }}{{X}^{2}}\frac{\sin \left( 1-\alpha  \right)\pi }{\sin \left( \frac{1-\alpha }{2} \right)\pi }.
\end{equation}
\item Now we calculate the mean energy loss due to the viscous damping force in the classical oscillator. From the relation(28), we have
\[J\left( t \right)={{c}_{meq}}{{\dot{x}}^{2}}\left( t \right),\]
where ${{c}_{meq}}$is denoted as the mean equivalent viscous damping coefficient for the fractional viscoelastic damping.

Then the mean of the energy loss is derived as
\begin{equation}
\overline{J}\left( t \right)={{c}_{meq}}\overline{{{{\dot{x}}}^{2}}}=\frac{1}{2}{{c}_{meq}}\dot{x}{{\dot{x}}^{*}}=\frac{1}{2}{{c}_{meq}}{{\Omega }^{2}}{{X}^{2}}.
\end{equation}

\item Letting $\overline{{{J}_{\alpha }}}\left( t \right)=\overline{J}\left( t \right)$ and from the relations (33) and (34) one derives
    \[\frac{c}{4}{{\Omega }^{1+\alpha }}{{X}^{2}}\frac{1}{2}\frac{\sin \left( 1-\alpha  \right)\pi }{\sin \left( \frac{1-\alpha }{2} \right)\pi }=\frac{1}{2}{{c}_{meq}}{{\Omega }^{2}}{{X}^{2}}.\]
   Consequently, we obtain the mean equivalent viscous damping coefficient for the fractional viscoelastic damping
\begin{equation}
{{c}_{meq}}=\frac{c}{2}{{\Omega }^{\alpha -1}}\frac{\sin \left( 1-\alpha  \right)\pi }{\sin \left( \frac{1-\alpha }{2} \right)\pi }.
\end{equation}

\end{Notes}
It is clear from (35) that the mean equivalent viscous damping coefficient for the fractional viscoelastic damping is a function of the vibration frequency $\Omega$ and the order $\alpha $ of the fractional derivative.
To this point, the fractional differential equations for the SDOF fractional oscillator (8) is approximately simplified to the following classical ordinary differential equation
\begin{equation}
m\ddot{x}\left( t \right)+{{c}_{meq}}\dot{x}\left( t \right)+kx\left( t \right)=f\left( t \right).
\end{equation}
With the aid of numerical simulations, we compare the vibration responses of the approximate integer-order oscillator (36) with the fractional one (8). The coefficients are respectively taken as $m=1$, $c=0.4$, $k=2$, $\alpha =0.56$, the external force is taken as the form $f=F\cos \Omega t$, where $F=30$, $\Omega =6$. In terms of Eq.(35), we derive the mean equivalent viscous damping coefficient ${{c}_{meq}}=0.14$.
\begin{figure}[H]
\centering
\includegraphics[scale=0.8]{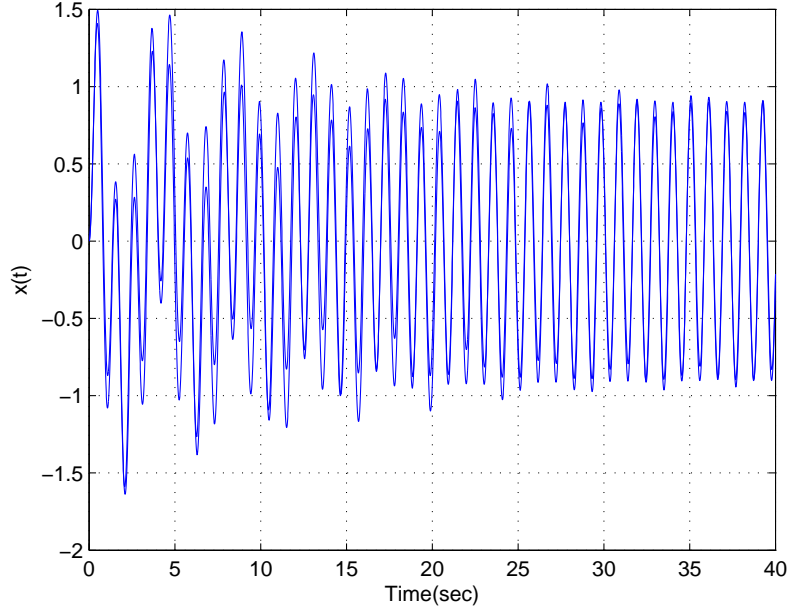}
\caption{The mean equilvalent damping coefficient for the SDOF fractional oscillator of Kelvin-Voigt type.}
\end{figure}
\begin{rmka}
By the above procedure, we can furthermore evaluate the mean equivalent viscous damping coefficient for the SDOF fractional oscillator (20) containing two fractional viscoelastic damping elements. Letting ${{J}_{{{\alpha }_{1}}}}\left( t \right)+{{J}_{\alpha 2}}\left( t \right)=\overline{J}\left( t \right)$ and from the relations (24) (25) and (34), we get
\begin{equation}
\begin{split}
& \quad\frac{{{c}_{1}}}{4}{{\Omega }^{{{\alpha }_{1}}+1}}{{X}^{2}}\frac{\sin \left( 1-{{\alpha }_{1}} \right)\pi }{\sin \left( \frac{1-{{\alpha }_{1}}}{2} \right)\pi }+\frac{{{c}_{2}}}{4}{{\Omega }^{{{\alpha }_{2}}+1}}{{X}^{2}}\frac{\sin \left( 1-{{\alpha }_{2}} \right)\pi }{\sin \left( \frac{1-{{\alpha }_{2}}}{2} \right)\pi }\\
&=\frac{1}{2}c\left( {{\alpha }_{1}},{{\alpha }_{2}},\Omega  \right){{\Omega }^{2}}{{X}^{2}}.\\
\end{split}
\end{equation}
From (37) we obtain the mean equivalent viscous damping coefficient
\begin{equation}
{{c}_{meq}}=\frac{{{c}_{1}}}{2}{{\Omega }^{{{\alpha }_{1}}-1}}\frac{\sin \left( 1-{{\alpha }_{1}} \right)\pi }{\sin \left( \frac{1-{{\alpha }_{1}}}{2} \right)\pi }+\frac{{{c}_{2}}}{2}{{\Omega }^{{{\alpha }_{2}}-1}}\frac{\sin \left( 1-{{\alpha }_{2}} \right)\pi }{\sin \left( \frac{1-{{\alpha }_{2}}}{2} \right)\pi }.
\end{equation}
With the aid of numerical simulations, we compare the vibration responses of the approximate integer-order oscillator (36) with the fractional one(20). The coefficients are respectively taken as $m=1$, ${{c}_{1}}=0.4$, ${{c}_{2}}=0.2$, $k=2$, ${{\alpha }_{1}}=0.56$, ${{\alpha }_{2}}=0.2$, the external force is taken as the form $f=F\cos \Omega t$, where $F=30$, $\Omega =6$. In terms of Eq.(38), we derive the mean equivalent viscous damping coefficient  ${{c}_{meq}}=0.56$.
\begin{figure}[H]
\centering
\includegraphics[scale=0.8]{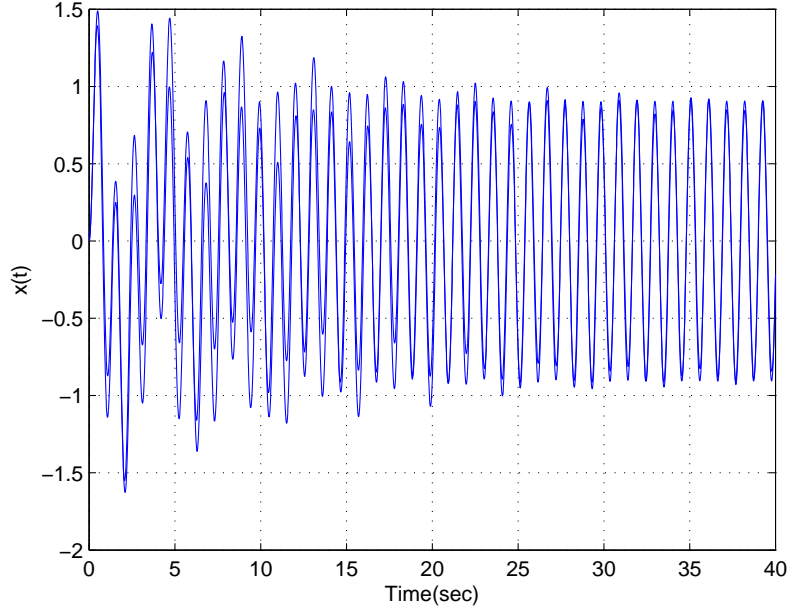}
\caption{The mean equilvalent damping coefficient for the SDOF fractional oscillator of modified Kelvin-Voigt type.}
\end{figure}
\end{rmka}
\section{Discussion}
The total mechanical energy in single degree of freedom fractional oscillators has been dealt with in this paper. Based on the energy storage and dissipation properties of  the Caputo fractional derivative, the total mechanical energy is expressed as the sum of the kinetic energy of the mass $\frac{1}{2}m{{\dot{x}}^{2}}$, the potential energy corresponding to the spring element $\frac{1}{2}k{{x}^{2}}$, and the potential energy stored in the fractional derivative $e\left( t \right)=\frac{1}{2}\int_{0}^{\infty }{{{\mu }_{1-\alpha }}\left( \omega  \right){{z}^{2}}\left( \omega ,t \right)d\omega }$. The energy regeneration and loss in vibratory motion have been analyzed by means of the total mechanical energy. Furthermore, based on the mean energy dissipation of the fractional damping element in steady-state vibration, a new concept of mean equivalent viscous damping has been suggested and the expression of the damping coefficient has been evaluated.

By virtue of the total mechanical energy in SDOF fractional oscillators, it becomes possible to formulate Lyapunov functions for stability analysis and control design for fractionally damped systems as well as other types of fractional dynamic systems. As for the future perspectives, our research efforts will be focused on fractional control design for fractionally damped oscillators and structures.

\textbf{Acknowledgements}

The author Yuan Jian expresses his thanks to Prof. Dong Kehai from Naval Aeronautical and Astronautical University, and Prof. Jiang Jianping from national University of Defense technology. All the authors acknowledge the valuable suggestions from the peer reviewers. This work was supported by the Natural Science Foundation of the Province Shandong of China titled Controls for fractional systems with applications to hypersonic vehicles (Grant Nos. ZR2014AM006).

\end{document}